\documentclass[conference]{IEEEtran}

\usepackage{cite}
\usepackage{amsmath,amssymb,amsfonts}
\usepackage{algorithmic}
\usepackage{graphicx}
\usepackage{textcomp}
\usepackage{xcolor}
\def\BibTeX{{\rm B\kern-.05em{\sc i\kern-.025em b}\kern-.08em
    T\kern-.1667em\lower.7ex\hbox{E}\kern-.125emX}}
    
\usepackage{tabularx}
\usepackage{booktabs}
\usepackage{calc}
\usepackage{enumitem}
\usepackage{url}
\usepackage{tikz}
\usepackage{xcolor}
\usepackage{balance}

\definecolor{dblue}{HTML}{1f77b4}
\definecolor{lblue}{HTML}{aec7e8}
\definecolor{dorange}{HTML}{ff7f0e}
\definecolor{lorange}{HTML}{ffbb78}
\definecolor{dgreen}{HTML}{2ca02c}
\definecolor{lgreen}{HTML}{98df8a}
\definecolor{dred}{HTML}{d62728}
\definecolor{lred}{HTML}{ff9896}
\definecolor{dviolet}{HTML}{9467bd}
\definecolor{lviolet}{HTML}{c5b0d5}
\definecolor{dbrown}{HTML}{8c564b}

\newcommand{\mybar}[2]{%
	\textcolor{black}{\rule{0.001pt*\real{#1}}{1.5ex}}\hskip5pt\hfill\textcolor{black}{{\footnotesize #2}}
}
\newcommand{\mybart}[2]{%
	\textcolor{black}{\rule{0.006pt*\real{#1}}{1.5ex}}\hskip5pt\hfill\textcolor{black}{{\footnotesize #2}}
}

\newcommand{\ccircle}[1]{%
    \tikz\draw[#1,fill=#1] (0,0) circle (.65ex);
}

\newcommand{\tweet}[1]{%
    \begin{center}
        \texttt{\small #1}
    \end{center}
}
    
\begin{document}

\title{Tweeting your \emph{Destiny}: Profiling Users in the Twitter Landscape around an Online Game}

\author{\IEEEauthorblockN{G\"unter Wallner\IEEEauthorrefmark{1}, Simone Kriglstein\IEEEauthorrefmark{2} and Anders Drachen\IEEEauthorrefmark{3}}
\IEEEauthorblockA{\IEEEauthorrefmark{1}Institute of Art \& Technology, University of Applied Arts Vienna, Vienna Austria\\
Email: guenter.wallner@uni-ak.ac.at}
\IEEEauthorblockA{\IEEEauthorrefmark{2}Center for Technology Experience, Austrian Institute of Technology, Vienna, Austria\\
Email: simone.kriglstein@ait.ac.at}
\IEEEauthorblockA{\IEEEauthorrefmark{3}DC Labs, University of York, York, United Kingdom\\
Email: anders.drachen@york.ac.uk}
}

\IEEEoverridecommandlockouts
\IEEEpubid{\begin{minipage}{\textwidth}\ \\[12pt]
978-1-7281-1884-0/19/\$31.00 \copyright 2019 IEEE \end{minipage}}

\maketitle

\begin{abstract}
Social media has become a major communication channel for communities centered around video games. Consequently, social media offers a rich data source to study online communities and the discussions evolving around games. Towards this end, we explore a large-scale dataset consisting of over 1 million tweets related to the online multiplayer shooter \emph{Destiny} and spanning a time period of about 14 months using unsupervised clustering and topic modelling. Furthermore, we correlate Twitter activity of over 3,000 players with their playtime. Our results contribute to the understanding of online player communities by identifying distinct player groups with respect to their Twitter characteristics, describing subgroups within the \emph{Destiny} community, and uncovering broad topics of community interest.
\end{abstract}

\begin{IEEEkeywords}
Twitter, topic modelling, community analysis, Destiny, MMOG, profiling, game analytics
\end{IEEEkeywords}

\section{Introduction}

Building healthy player communities has long been recognized as an important factor for the market success of online games (cf.~\cite{Ruggles:2002}). Studying the social connections within online games has thus seen a growing interest over the years (see, e.g.,~\cite{Warmelink:2011}). However, social activity has long escaped the boundaries of the games themselves and game-related discourse is now taking place across all sorts of online discussion platforms. Indeed, gamers constitute an active subgroup of social media users (cf.~\cite{Kalaitzis:2016}) making social media, including micro-blogging services such as Twitter, a valuable source for studying gamer communities. Despite being a rich data source, research using Twitter to understand the public discussion around games is still relatively sparse (see Section~\ref{sec:rw} for some notable exceptions). In our own previous work~\cite{Drescher:2018} we proposed a visual analytics system and demonstrated the usefulness of such an approach for qualitatively analyzing tweets in the context of games user research using a large-scale dataset of tweets related to the game \emph{Destiny}~\cite{Destiny:2014}. \emph{Destiny} is a multiplayer first-person shooter set in a science fiction world where players compete against each other in different player vs player (PvP) and player vs environment (PvE) game modes. It is a suitable use case for this kind of analysis because it exposes gameplay data through a public API and was supported by the developer over an extended time period. In this paper, relying on the same dataset, we continue our efforts in exploring community engagement on Twitter by taking a quantitative approach and utilizing unsupervised techniques, specifically clustering and topic modelling by means of Latent Dirichlet allocation (LDA), for the purpose of:

\begin{itemize}
    \item Assessing the relationship between, and establishing player profiles with respect to, tweeting behaviour and playtime;
    \item Uncovering subgroups within the \emph{Destiny} community based on how people identify themselves in their public Twitter profile descriptions;
    \item Identifying broad topics of interest to the community, their variation over time, and if and how these interests differ between subgroups.
\end{itemize}

\noindent In that sense, this paper contributes to creating a better understanding of online communities which form around games. This can help to maintain and shape a strong player community and, in turn, extend the longevity of a game. Insights gained through the proposed analysis methodologies can point to new directions for analyzing motivations and to specific preferences of players.

\IEEEpubidadjcol
\section{Related Work}
    \label{sec:rw}

Player communities have been studied not only within games (see~\cite{Warmelink:2011} for a good overview) but also increasingly outside of games. Due to the focus of this paper we will restrict the discussion to the latter category and on work employing quantitative methods for analyzing Twitter datasets. For example, McDonald and Moffat~\cite{McDonald:2016} conducted a sentiment analysis to gain insights into the positive and negative reactions to the yearly changing theme of the Global Game Jam by capturing and analyzing Twitter data. Chatzakou et al.~\cite{Chatzakou:2017} undertook a quantitative analysis of tweets in relation to the Gamergate controversy. For this purpose, a dataset of 340,000 users and 1.6 million tweets was analyzed to understand users' properties and to investigate the content they posted as well if these were different from random Twitter users. Kalaitzis et al.~\cite{Kalaitzis:2016} presented a method for predicting gaming-related properties from Twitter profiles by applying machine learning models on labelled accounts. 

In our own previous work~\cite{Drescher:2018}, we used visual analytics methods to identify which topics matter to \emph{Destiny} players, to find influential members of the game's community, and to untangle the relationship between contextual information about the tweets and in-game activity. However, no automated topic modelling was performed. In contrast, in the present paper we make use of topic modelling to extract key themes from a large corpus of tweets. Topic modelling has been used extensively in other domains (e.g.,~\cite{nigam2017,prier2011,surian2016}) but, to our best knowledge, there exist only a few works within in the context of games research. These are usually focused on the analysis of chat logs but not Twitter data. For example, Tinati et al.~\cite{Tinati:2015} analyzed over two years of gaming and chat data, comprised of over 4 million completed game sessions and 885,000 chat entries, from the citizen science game \emph{EyeWire} to better understand how features of the game contribute to player communication. To examine the content of the chat messages topic modelling was performed. This helped them to identify the vocabulary as well as the discourse that occurred during the different stages of the game. Musabirov et al.~\cite{Musabirov:2015} used text mining techniques, including topic modelling on game chat logs to improve tutorials in massively multiplayer online games. In order to investigate how the gender of streamers is associated with the nature of conversation, Nakandala et al.~\cite{Nakandala:2017} analyzed chat messages from Twitch. Ryan et al.~\cite{Ryan:2015} presented GameNet, a tool for game discovery build upon a latent semantic analysis of Wikipedia articles for nearly 12,000 digital games to establish the semantic relatedness between them. The article also offers a comprehensive review of the use of natural language processing techniques in the context of game studies.

\section{Data Set}

The analysis presented in this paper is based on a dataset used in previous work~\cite{Drescher:2018} and which contains 1,062,230 tweets with relation to \emph{Destiny} within a time period of about 14 months (from June 9th, 2016 to July 17th, 2017) and details of 381,076 accounts tweeting about \emph{Destiny}. It was obtained using the Twitter Search API~\cite{TwitterAPI} and the following query:  

\begin{center}
    \tt\footnotesize
    (destiny AND (e3 OR activision OR bungie OR crucible OR crota OR guardians OR ironbanner OR psn OR xbl OR gamertag OR gameplay)) OR riseofiron OR destinygame OR destinythegame OR gjallarhorn
\end{center}

The search term \texttt{Destiny} was combined with different game-related key words to reduce the odds of collecting tweets unrelated to the game. Besides the corpus of tweets, the dataset includes \emph{Destiny} in-game behavioral data over a six-month period (June 10th, 2016 to December 1st, 2016) from 3,547 players who shared their PSN ID or Xbox Gamertag in their Twitter profile description. These IDs were extracted from the profile description using various regular expressions (cf.~\cite{Drescher:2018}).

While the dataset contains a variety of in-game behavioral features, we will focus on a subset of those which are related to the overall time played during the above period, that is, time spent playing in the various PvP and PvE modes that \emph{Destiny} offers. This information was gathered through the Bungie API~\cite{BungieAPI:2017} using the gamertags and PSN IDs provided in the profile descriptions. This allows us to assess if gameplay duration is connected to tweeting behaviour. 

\section{Tweet Behaviour}

To asses if and how activity on Twitter corresponds with in-game activity (i.e., time spend playing) we first focus on the 3,547 Twitter accounts for which in-game data is available. Specifically, we extracted the total time spend playing (PvP and PvE combined) and measures of Twitter activity (number of tweets, retweets, and replies as well as the total number of people a user is following) and indicators of popularity (number of followers, average number of retweets a tweet receives, average number of likes a tweet receives) in order to correlate the measures and to develop player profiles through $k$-means clustering. 

\subsection{Data Preprocessing}

Of the 3,547 accounts for which gameplay data is available, 114 did not tweet during the six-month period and were thus excluded from analysis. Since $k$-means clustering is known to be sensitive to outliers (cf.~\cite{Han:2011}) and because some players exhibited very large values along one of the features (e.g., comparatively very high number of followers or play duration), we also removed entries with a $z$-score greater than $\pm 3$ (i.e. removing the outlying 0.15\% of both tails of the frequency distribution) in at least one column. 

Of the 314 accounts removed, 253 formed an outlier among one feature and 61 formed an outlier on two or three features. The main contributor to outliers was the average number of retweets with 74 outliers, followed by the number of tweets with 69 outliers. However, all measures had outliers removed. The outliers thus mainly represent Twitter users who were exceptionally active or inactive along one dimension. This process left 3,119 accounts for analysis. Finally, given the different scales of the various metrics, we used min-max normalization to scale each metric to the range $[0...1]$. 

\subsection{Analysis and Results}

\begin{figure*}
    \centering
    \includegraphics[width=0.9\linewidth]{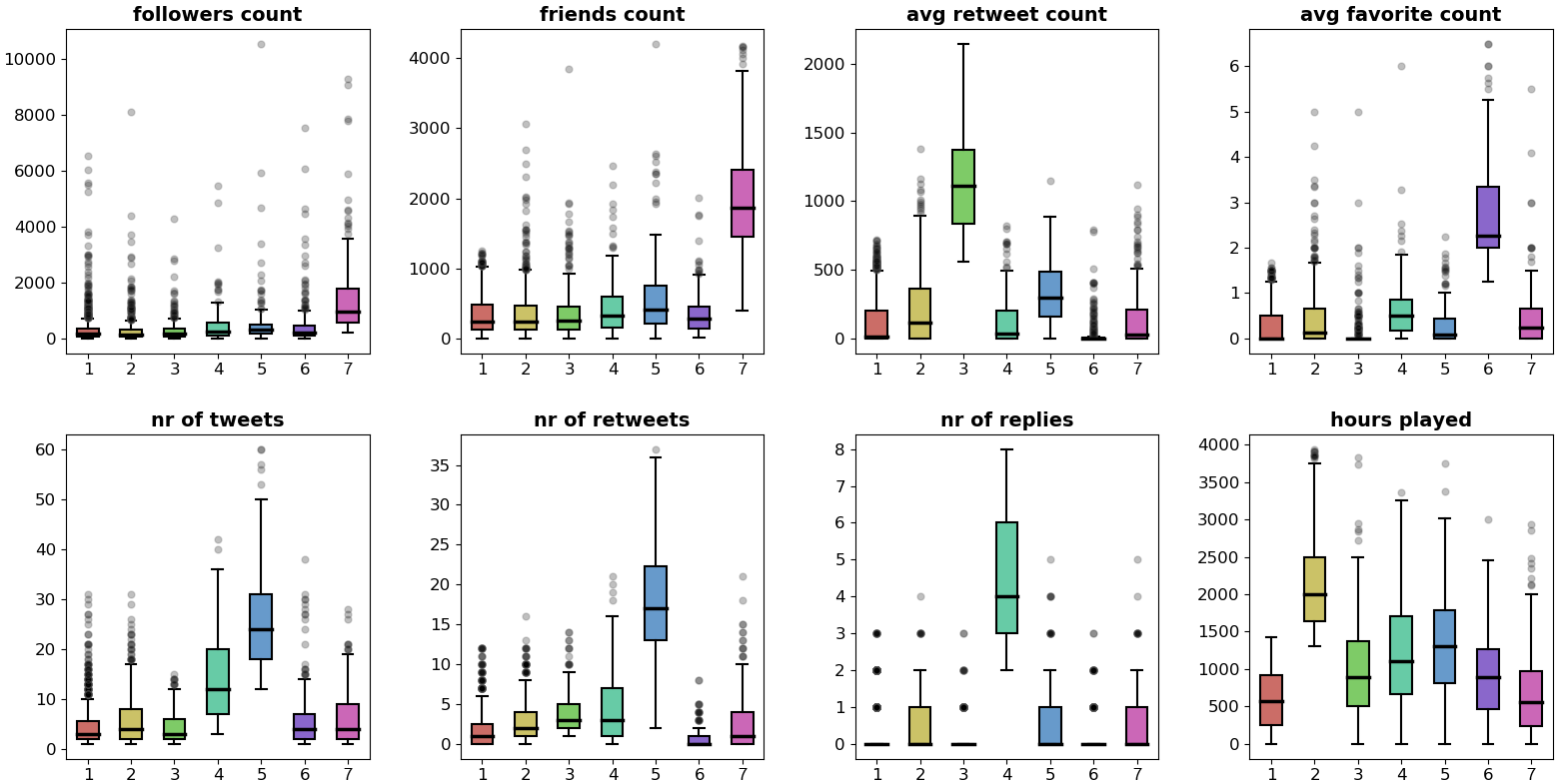}
    \caption{Variation of various Twitter related features and play duration across the seven identified clusters.}
    \label{fig:clusters}
\end{figure*}

A Spearman rank correlation did only yield weak correlations between the Twitter related measures and in-game activity with correlation coefficients between $-0.15$ and $0.18$. Using scikit-learn~\cite{scikit-learn}, $k$-means clustering was performed with $k = 2...15$ and the resulting clusterings were assessed using the silhouette coefficient, pointing to a seven-cluster solution as described in the following. Figure~\ref{fig:clusters} shows how the included measures vary between the seven identified profiles.

\begin{description}[leftmargin=0cm]
    \item[Profile~1 - Minimally Active:] This cluster constitutes the largest group covering $~$43\% of the whole sample ($N= 1,343$) and is comprised of people who show low activity on Twitter, have a small number of followers, and have (together with Profile~7) the lowest duration of play.
    \item[Profile~2 - Highly Active Gamers:] This cluster includes Twitter users ($N = 644$) who spend considerable time playing \emph{Destiny}, in fact their play duration is the highest across all profiles. These can be considered highly active gamers but at the same time they do not engage much on Twitter, with tweeting related measures being similar to the average user of Profile~1. 
    \item[Profile~3 - Influencers:] Tweets of players in this group ($N$~=~320) are often shared as they receive on average a large number of retweets. While they show less tweeting activity than players in Profile~5 they contribute to information diffusion to a much larger extent. 
    \item[Profile~4 - Respondants:] Players in this group ($N = 149$) are also quite active on Twitter in terms of the number of tweets and retweets (but less so than players belonging to Profile~5). However, they are primarily characterized by making a large number of replies compared to other profiles. 
    \item[Profile~5 - Socially Active:] This group of players ($N = 192$) is not only very active in the game (but considerable less than players belonging to Profile~1) but are also highly active on Twitter: they tweet a lot about \emph{Destiny} and make many retweets. At the same time and despite being highly active on Twitter they do not attract as many retweets on average as players belonging to Profile~3. 
    \item[Profile~6 - Favourites:] Tweets from players in this group receive on average the most likes (avg. favorite count) but at the same time receive the lowest average number of retweets across all profiles ($N = 265$). 
    \item[Profile~7 - Followers:] Players in this group ($N = 206$) show -- together with players of Profile~1 -- the least in-game activity across the identified clusters. Similarly, they are also among the less active users on Twitter. What sets them apart from other profiles is that they follow a considerable larger number of people (friend count) and also have a larger number of followers than players belonging to the other profiles. These seem to be players that act as 'information distributors' which follow other people to gather information and share it with there followers. These tweets are, however, considered less worthy of retweeting.
\end{description}

\section{Topic Modelling}

Topic modelling is the process of finding topics (i.e., groups of words) which frequently occur in a collection of documents. For our analysis we have opted for Latent Dirichlet allocation (LDA)~\cite{Blei:2003} as it is widely used in natural language processing and produces topics that are well interpretable by humans (cf.~\cite{Chang:2009}). LDA is a probabilistic model, modelling each document as a mixture of latent topics, with the topics being represented by a distribution over words. In other words, LDA takes as input a collection of documents assumed to contain $k$ topics. Each document belongs to the $k$ identified topics with differing probabilities. 

\begin{table*}[t]
    \setlength{\tabcolsep}{5.5pt}
    \caption{Topics obtained through LDA analysis of profile descriptions.}
    \label{table:ldaTopicsProfiles}
    \begin{tabularx}{\textwidth}{llll@{\hskip5pt}p{2.7cm}}
        \toprule
        \textbf{Label} & \textbf{$\mathbf{c_v}$} & \textbf{Top Keywords}$^\dagger$ & \multicolumn{2}{l}{\textbf{No. and \% of Profiles}} \\
        \midrule
    Streamers & 0.579 & twitch\_streamer, youtube, youtube\_channel, twitch, year\_old, streamer, channel                   & 25,245 & \mybar{25245}{11.3\%} \\ 
    Gaming Enthusiasts & 0.516 & xbox\_one, destiny, call\_duty, xbox, psn, ps4, follow, follow\_back, one, player            & 19,179 & \mybar{19179}{8.6\%}\\
    Hobby Streamers & 0.442 & full\_time, part\_time, time, come\_check, let\_play, full, rainbow\_six, part               & 16,228 & \mybar{16228}{7.3\%} \\
    Gamers & 0.430 & video\_game, game, video, love, like, play, life, thing, make, gaming                     & 44,842 & \mybar{44842}{20.1\%}\\
    News Outlets & 0.420 & gaming\_news, gaming, news, subscribe\_youtube, social\_medium, gaming\_community               & 22,991 & \mybar{22991}{10.3\%}\\
    Students \& SciFi Fans & 0.390 & web\_chat, star\_war, college\_student, cod\_player, add\_xbox, come\_join, star                    & 14,696 & \mybar{14696}{6.6\%}\\
    Casual Gamers & 0.382 &  gamer, video\_game, fan, lover, music, game, husband\_father, sport, gaming                    & 32,361 & \mybar{32361}{14.5\%}\\
    Creative People & 0.347 & destiny\_player, content\_creator, graphic\_designer, like\_body, live\_chat, coming\_soon   & 15,621 & \mybar{15621}{7.0\%}\\
    Official Accounts &  0.345 &official\_twitter, twitter, official, business\_inquiry, twitter\_account, playing\_video  & 15,409 & \mybar{15409}{6.9\%}\\
    Clans \& Members & 0.280 & proud\_member, check\_channel, destiny\_clan, one\_day, make\_sure, member                  & 16,887 & \mybar{16887}{7.6\%}\\
        \bottomrule
    \end{tabularx}
    \begin{flushright}
        \vskip-3pt
        $^\dagger$Keywords are sorted by decreasing probability.\hfill Words connected by underscores represent bigrams. 
    \end{flushright}
    \vskip-1.0\baselineskip
\end{table*}

\subsection{Data Preprocessing}
We first filtered out all accounts with empty or non-English profile descriptions. For the latter, we followed the recommendation of Lui and Baldwin~\cite{Lui:2014} and used a majority vote across the three language detection algorithms \texttt{langid}~\cite{Lui:2012}, \texttt{langDetect}~\cite{Nakatani:2010}, and \texttt{CLD2}~\cite{Al-Rfou:2015}. That is, only accounts where at least two of the three algorithms classified the description as English were retained. Next, we removed fake accounts such as multiple accounts with exactly the same profile description or where to description was clearly suspicious (e.g., several accounts claiming to give away smartphones). In case of tweets, we dropped non-English tweets (following the procedure above), duplicate tweets, tweets with exactly the same text but including different URLs, tweets from the fake accounts identified before, and retweets. This resulted in 352,088 tweets and 223,459 accounts for topic modelling. 

Following established procedures, tweet texts and profile descriptions were tokenized. Subsequently, all tokens were converted to lower case and punctuation characters were removed. Tokens representing numbers, mentions of other accounts (i.e., tokens starting with an $@$), and URLs were removed as were tokens with less than three characters and tokens representing stopwords. All remaining tokens were lemmatized to obtain their base form and to reduce data sparseness using the WordNet lemmatizer provided by NLTK~\cite{Bird:2009}. Finally, tokens appearing in more than 80\% of the documents as well as tokens only occurring in less than 3 documents were removed.

\begin{table*}[t]
    \caption{Prevalent topics in tweets as identified by LDA analysis.}
    \label{table:ldaTopicsTweets}
    \setlength{\tabcolsep}{4pt}
    \begin{tabularx}{\textwidth}{llll@{\hskip5pt}p{2.5cm}}
        \toprule
        \textbf{Label} & \textbf{$\mathbf{c_v}$} & \textbf{Top Keywords}$^\dagger$ & \multicolumn{2}{l}{\textbf{No. and \% of Tweets}} \\
        \midrule
        Iron Banner      &  0.463 & iron\_banner, banner, iron, ironbanner, supremacy, control, surprised, destiny, gamestop    & 5,785 & \mybart{5785}{1.6\%} \\
        Rise of Iron     &  0.438 & rise\_iron, rise, iron, bro, e32017, dlc, destiny, via, riseofiron, gameplay                & 7,782 & \mybart{7782}{2.2\%} \\
        Trials of Osiris &  0.503 & trial\_osiris, trialsofosiris, last\_night, trial, add, night, flawless, osiris, add\_psn   & 7,385 & \mybart{7385}{2.1\%} \\
        Live Streaming   &  0.470 & live, supportsmallstreamers, twitch, destinythegame, playing, stream, morning, destiny      & 8,019 & \mybart{8019}{2.3\%} \\
        Announcing Videos  & 0.510 &   added\_video, added, playlist, video, gameplay, basically, episode, story\_mission, podcast & 4,413 & \mybart{4413}{1.3\%} \\
        Liking Videos    & 0.394 & liked\_video, liked, video, gameplay, really\_hope, fam, agree, destiny, crucible, hope     & 4,906 & \mybart{4906}{1.4\%} \\
        Announcements &  0.380 & release\_date, gameplay\_trailer, release, date, trailer, sorry, seems, gameplay, called    & 4,684 & \mybart{4684}{1.3\%} \\
        Gameplay &  0.320 &  gameplay\_reveal, reveal, gameplay, cayde, via, destiny, grinding, idk, watch, red\_bar     & 5,492 & \mybart{5492}{1.6\%} \\
        Infrastructure &  0.320 &  server, destiny, wtf, riseofiron, ps4live, queue, happened, tapir, destinyriseofiron        & 6,225 & \mybart{6225}{1.8\%} \\
        (Dis)Likes/Observations &  0.381 & look\_like, look, feel\_like, like, gary, next\_week, feel, holy\_shit, birthday, holy      & 6,718 & \mybart{6718}{1.9\%} \\
        Various &  0.327 & life, boy, reason, bought, seen, mine, hoping, crucible\_highlight, gave, good\_luck        & 4,208 & \mybart{4208}{1.2\%} \\
        \bottomrule
    \end{tabularx}
    \begin{flushright}
        \vskip-3pt
        $^\dagger$Keywords are sorted by decreasing probability.\hfill Words connected by underscores represent bigrams.
    \end{flushright}
    \vskip-1.0\baselineskip
\end{table*}

\begin{figure*}
    \centering
    \includegraphics[width=1.0\linewidth]{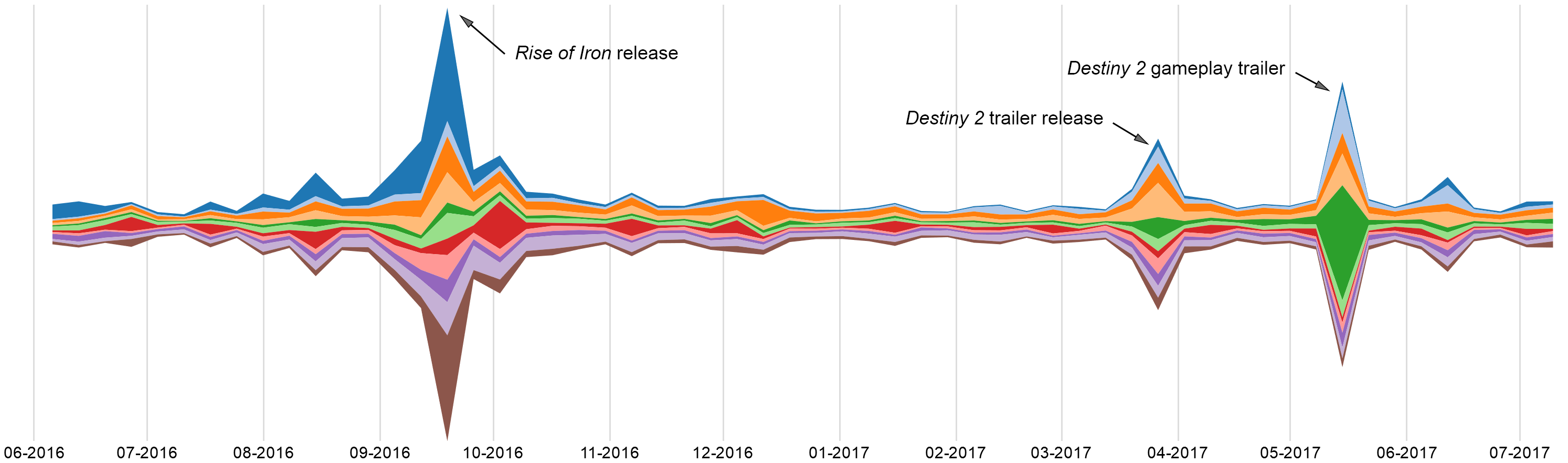}
    \caption{Streamgraph of the 11 most coherent and frequent topics over time (June 9th, 2016 to July 17th, 2017, \protect\ccircle{dblue}Rise of Iron, \protect\ccircle{lblue}Announcements, \protect\ccircle{dorange}Live Streaming, \protect\ccircle{lorange}(Dis)Likes/Observations, \protect\ccircle{dgreen}Gameplay, \protect\ccircle{lgreen}Liking Videos, \protect\ccircle{dred}Iron Banner, \protect\ccircle{lred}Various, \protect\ccircle{dviolet}Announcing Videos, \protect\ccircle{lviolet}Trials of Osiris, \protect\ccircle{dbrown}Infrastructure).}
    \label{fig:streamgraph}
\end{figure*}

\subsection{Analysis}

We performed LDA on both the corpus of tokenized and lemmatized profile descriptions and tweet texts (extended with bigrams and trigrams of the tokens) using the Python library \texttt{gensim}~\cite{Rehurek:2010}. A TF-IDF~\cite{Salton:1986} term-weighting bag-of-words model was used to adjust for the fact that some words occur more frequently across documents (e.g., \emph{Destiny} in our case). LDA models where calculated from 2 to 20 topics for profile descriptions and 2 to 200 topics for tweet texts. Lastly, each profile description and tweet was assigned to the most probably topic.

To rate the quality of the topics, we assessed the coherence of each model using the $c_V$ coherence score proposed by Roder et al.~\cite{Roder:2015} and which has been shown by the authors to align well with human judgment. After inspection of the resulting plots and inspection of the topics for interpretability we opted for a 10 topic solution for the profile descriptions. In case of tweets, coherence did not show a distinctive elbow but increased slowly until around 175 topics after which it flattened out. For that reason, we also estimated the perplexity of each topic model showing a steep decline between 76 and 84 topics, thus either pointing to a 84 or 175 topic solution. While such a large number of topics can be expected (assuming that a broad range of issues is discussed by the community) it is also known that -- in such a case -- LDA tends to produce many 'junk' topics  (cf.~\cite{Nikolenko:2017,Wang:2016}), that is, topics which do not capture the underlying semantics well and are thus hard to interpret by humans. After manual inspection of both models we found that the 175 solutions does not add much information to the 84 solution which itself already contains topics that are difficult to interpret. To keep the following analysis focused on the more common and interpretable topics we thus only retain those with a topic coherence score greater than the average ($>$ 0.304) and which encompass a higher than average number of tweets ($>$ 4,191).
 
\subsection{Results}

Table~\ref{table:ldaTopicsProfiles} lists the profile topics together with the topic coherence score $c_v$, a list of most likely keywords, as well as the number and percentage of profiles belonging to each topic. Bi-grams are represented using an underscore. Similarly, Table~\ref{table:ldaTopicsTweets} lists the same information for topics inferred from the tweets. Focusing first on the profile descriptions we could derive the following topics on how people identify themselves on Twitter:

\begin{description}[leftmargin=0cm]
    \item[Gaming Enthusiasts:] This group frequently and predominantly uses gaming-related terms such as Xbox and Xbox One, PSN, PS4, and different games such as \emph{Call of Duty} to describe themselves on Twitter. For this reason we have coined them gaming enthusiasts.  
    \item[Gamers:] These constitute the largest group within the \emph{Destiny} audience, expressing their love for video games or how they like playing video games in their profile descriptions. These are people who we consider avid gamers but who use less specific terms or games than \textit{Gaming Enthusiasts} to express their interest.
    \item[Casual Gamers:] These are people who do not identify themselves only as gamers in their profiles but for which gaming is one hobby next to, for example, music or sports. 
    \item[Clans \& Members: ] Given the multiplayer focus of \emph{Destiny} and the clan culture surrounding the game it is interesting to explicitly observe a number of Twitter accounts which, for example, belong to clans or to people who identify themselves as members of a clan. 
    \item[Students \& SciFi Fans: ] People identifying first and foremost as students (but with a certain gaming affinity) and/or science fiction fans (e.g., \emph{Star Wars}) constitute another group within the \emph{Destiny} community. Given that \emph{Destiny} takes place in a science-fiction setting it is fitting to see that the game resonates with the latter group. 
    \item[Streamers: ] Within this topic fall players who describe themselves as streamers (e.g., Twitch) or who are maintaining a gaming-related channel, for example, on YouTube.
    \item[Hobby Streamers: ] This group is related to the above as people also express their interest in streaming, but are not as engaged in it judging from the keywords (e.g., part-time). We thus labelled this group as hobbyists.
    \item[News Outlets: ] These accounts are related to some sort of news outlets such as, for example, gaming websites, magazines, or YouTube channels that cover gaming-related news.
    \item[Creative People: ] \emph{Destiny} seems to also attract -- what we coined -- creative people, that is, people such as graphic designers or content creators which also highlight their interest in the game in their public profile description.
    \item[Official Accounts: ] Profile descriptions within this topic point to various official accounts or accounts of businesses which tweet about \emph{Destiny} or where, for example, the business is related to gaming.     
\end{description}

\noindent Shifting the focus to topics identified in the tweets themselves (cf. Table~\ref{table:ldaTopicsTweets}) we witness a number of topics pertaining to in-game events, including the weekly \emph{Trials of Osiris} event and the once-per-month held \emph{Iron Banner} as well as to game extensions such as \emph{Rise of Iron} (released on September 20th, 2016). For example:

\tweet{Time for some iron banner. Let's play!!!  \#xbox \#xboxone \#bungie \#destiny \#destinythegame \#RiseofIron\ldots} 

\noindent Other topics evolve around news or announcements regarding the game such as trailers, release dates, or gameplay related discussions such as the reveal of new gameplay material, e.g.:

\tweet{Only 14 days until the Destiny 2 gameplay reveal. I hope they show the Crucible, and any changes Bungie made!} 

\noindent Streaming and the creation and watching of videos are also prevalent topics, with people announcing that they are currently live streaming, as exemplified by the following tweet

\tweet{I'm live on Twitch - Watch me at  \#SupportSmallStreamers \#destiny \#riseofiron Clan raid and helps} 

\noindent or that they have uploaded new videos with others, in turn, expressing that they liked certain videos. Issues related to what we called \textit{Infrastructure} including server related issues, queuing for a match, or online platforms (e.g., PlayStation Live) form another common topic. 

People also frequently express what they like or do not like about the game or make observations, for instance:

\tweet{I was so hyped for destiny2, but all the gameplay just looks like another expansion of the old destiny. So might not even get it} 

\noindent Lastly, we find a topic (entitled \emph{Various} in Table~\ref{table:ldaTopicsTweets}) which is hard to grasp based on the keywords and assigned tweets and which covers quite a broad range of different issues such as people tweeting about how \emph{Destiny} is affecting their lives, such as:

\tweet{This is my life now. Good thing I don't have plans tomorrow. \#RiseOfIron \#destinythegame \#Destiny \#PS4share}

\noindent Figure~\ref{fig:streamgraph} shows how these topics vary over time. As evident from the figure, tweets surrounding the \emph{Rise of Iron} extension increase steadily until the release day but after about two to three weeks interest was low again. That is, while the extension seems to have been highly anticipated, interest seems to have faded quickly which might be in line with the short duration of the extension, ranging from only a couple of hours to about 20 hours.\footnote{\url{https://howlongtobeat.com/game.php?id=38186} (Accessed: March, 2019)} At the same time we can see a large increase of tweets related to \textit{Infrastructure} at the day of the \emph{Rise of Iron} release. Inspecting tweets falling within this topic on this day shows that the release was accompanied by server issues with people complaining about servers being down, for instance: 

\tweet{@DestinyTheGame Servers are down :\textasciicircum(}

\noindent Returning to topics related to in-game events, the graph also shows how conversation around the \emph{Iron Banner} is peaking once per month (at the time it is held). In terms of streaming we can see increased activity around the \emph{Rise of Iron} release but even more so in mid-December, 2016 around the time when \emph{The Dawning} went live on December, 13th. 

\tweet{Happy dawning guardians video and stream will be going on after I get off of work today!!! \#Destiny \#thedawning}

\noindent While \emph{The Dawning} in-game event lasted for about 3 weeks, tweets about streaming \emph{Destiny} where soon back to 'normal' a few days after it started. In contrast, the Halloween event \emph{Festival of the Lost} held in the last week of October and first week of November did not attract noticeable attention from streamers.

In relation to the \emph{Gameplay} topic, we can observe the largest activity in mid-May, 2017 (when a  \emph{Destiny 2} trailer with actual gameplay footage was released) with some increased activity since the announcement and release of the first \emph{Destiny 2} trailer about one and half months earlier (end of March, 2017). Announcements (a topic closely related to the previous one) are also increased during that time, peaking at the aforementioned two time periods. Expressing ones opinions (liking or disliking something) also strongly coincides with new releases (such as \emph{Rise of Iron}) or announcements (e.g., \emph{Destiny 2} trailer). That is, the community immediately expresses its thoughts and opinions with respect to new announcements. 

In terms of announcing and liking videos there is generally no strong variation over time but a somewhat increased activity can be witnessed after the \emph{Rise of Iron} release with people immediately starting to share new gameplay footage, among others walkthrough videos of the new content:

\tweet{I added a video to a @YouTube playlist \url{https://t.co/3iXEL8uu8Z} Destiny Rise of Iron Download Complete Walkthrough Gameplay Story}

\noindent The other topics show no notable variations during the analyzed time period.

\begin{figure}
    \centering
    \includegraphics[width=1.0\linewidth]{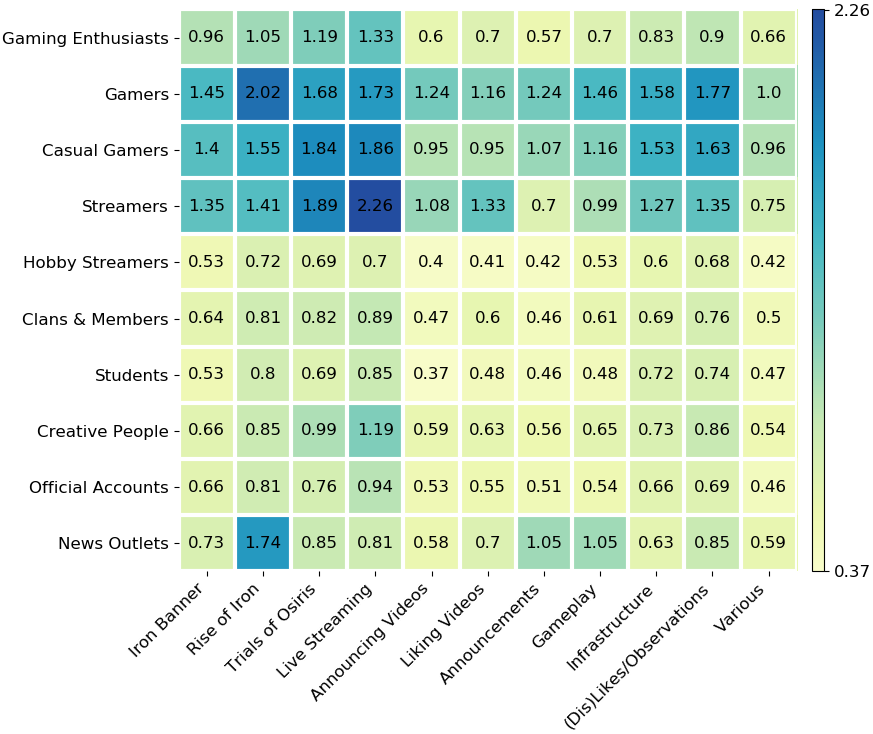}
    \caption{Tweet topic distribution ($x$-axis) across the 10 identified profile groups ($y$-axis) in percentages (60,469 tweets in total).}
    \label{fig:heatmap}
\end{figure}

Lastly, we investigated if and how the different identified groups differ in terms of which topics they are tweeting about. Figure~\ref{fig:heatmap} shows the percentage of tweets falling within each topic across the ten identified user groups. The most active groups with respect to the above tweet topics are \textit{Gaming Enthusiasts}, \textit{Casual Gamers}, and \textit{Streamers}. Streamers unsurprisingly tweet the most about streaming followed by different game-related events, with the \emph{Trials of Osiris} (a weekly event) garnering particular attention among streamers, more than \emph{Rise of Iron} or the \emph{Iron Banner}. The higher interest of streamers in the \emph{Trials of Osiris} could be a result of the more frequent conduction of the event, compared to the \emph{Iron Banner} which is only held once per month and \emph{Rise of Iron} for which -- as discussed above -- interest seemed to have faded quickly after its release. \textit{News Outlets} showed the highest interest in the \emph{Rise of Iron} extension (the major addition to the game in the observed time period), \textit{Announcements}, and \textit{Gameplay} material. \textit{Infrastructure} related issues are most important for \textit{Gamers} and \textit{Casual Gamers} but interestingly not so much for \textit{Gaming enthusiasts}. Otherwise we could not observe strong variations: \textit{Gamers} and \textit{Casual Gamers} show a similar pattern, as do \textit{Hobby Streamers}, \textit{Clans \& Members}, \textit{Students \& SciFi Fans}, \textit{Creative People}, and \textit{Official Accounts}.

\section{Discussion}

Focusing on a subset of Twitter users which shared their Xbox Gamertag or PSN ID we developed seven profiles of players based on their Twitter characteristics and in-game activity. While the largest fraction (Profile~1) shows only limited Twitter activity there is a much smaller group of very active players with respect to retweeting (Profile~5). These can be considered to mainly contribute to information spreading (perhaps because they are highly enthusiastic about the game judging from their duration of play) but they do not necessarily receive a lot of retweets themselves. While these are very active, there is another group of players (Profile~3) which can be considered more influential as their tweets receive more attention and are shared more widely. These are, however, not necessarily the ones with the most followers (which would be players belonging to Profile~7). This reflects the findings of Cha et al.~\cite{Cha:2010} who found that the number of followers is not related to the number of retweets, which is largely driven be the attributed content value. Besides there is also a group of players (Profile~6) who do not receive many retweets at all but at the same time garner a lot of likes. People appear to like the content these players post but at the same time deem it less worthy to distribute the information further. Others (Profile~4), in turn, seem to like to engage in discussions and replying to others. Particularly noticeable is that players spending the most time in \emph{Destiny} show comparable low Twitter activity (Profile~2). In a recent study by Canossa et al.~\cite{Canossa:2019}, influencers in in-game social networks (in terms of network centrality) showed only about one-fourth of the playtime of power players (i.e. the most engaged players). A similar phenomenon can be found in our study: the most engaged players (in terms of playtime) are not necessarily the most active or influential on Twitter.  

Using LDA we grouped the profile descriptions into ten reoccurring topics to get an impression of the demographics of the community. These topics can be grouped into larger categories: people identifying themselves as gamers to various degrees, people playing games but considering them first and foremost not as gamers, streamers, news outlets, and official accounts. It needs, however, to be acknowledged that the identified profiles are based on descriptions of how people see themselves, that is, someone using a lot of gaming related keywords may actually not be a professional gamer. That said, such a categorization can be helpful to profile the audience a game attracts and to offer content which matches the interest of specific subgroups. 

In terms of tweet content we identified several common topics as listed in Table~\ref{table:ldaTopicsTweets}. These topics are related to in-game content (e.g., events, extensions), streaming and video sharing, infrastructure such as server issues, expressing opinions, and news concerning the game. It has to be kept in mind that this list of topics is not exhaustive as we have focused on the most coherent and frequent topics. Comparing these topics with the themes identified by Drescher et al.~\cite{Drescher:2018} -- who used a visual analytics approach and did not rely on unsupervised topic modelling but on the same dataset -- we can observe some differences. In general, Drescher et al. identified more specific issues such as people looking for others to play with or people sharing walkthroughs. While LDA also produced a topic related to matches described by keywords such as private\_match, match, took, and private ($c_v = 0.326$, 4,179 tweets) the associated tweets covered all aspects related to matches such as players being excited about the introduction of private matches or posting about playing a match. Similarly, walkthroughs were not grouped into a single topic but were part of different topics. The topics produced by LDA should be viewed as broad themes, rather than encompassing specific aspects. In this sense, topic modelling allowed us to extract greater themes which can serve as a starting point to identify more specific issues. Further work may explore the use of (semi-)supervised methods, e.g., targeted topic modelling~\cite{Wang:2016}.

Plotting topics over time offers a means to study trends, to monitor audience interest and anticipation, and also to observe for how long the interest pertains. Quickly fading interest in, for example, events or extensions may provide an indication for when to release new content in order to sustain interest in a game. It is noticeable that interest vanishes quite quickly in many cases (e.g., with respect to new game content) which also affects streaming activity (as in case of \emph{The Dawning}). In light of this, regular events (e.g., \emph{Iron Banner}) may be a good way to keep the talk about a game going. Monitoring topics could also be an efficient means to detect issues and changes in audience attitudes and sentiment. 

When interpreting the results of this study some limitations need to be highlighted. First, due to the query that was used to retrieve the tweets some topics (e.g., \emph{Rise of Iron}) may be more prevalent than others. These additional constraints were added because if Twitter is only queried for \texttt{Destiny} it returns too many tweets totally unrelated to the game. If searched only for the official hashtag \texttt{\#DestinyTheGame}, the returned tweets only cover a small subset of \emph{Destiny} related tweets. That is, constructing an unbiased query which still ensures that the tweets relate to the game is a challenging task. In addition, the Twitter Search API used to construct the dataset is focused on relevance and not completeness.\footnote{\url{https://developer.twitter.com/en/docs/tweets/search/overview/standard.html} (Accessed: March 2019)} However, as the sampling is not detailed it is unclear if and which bias is introduced by the API. Please note, that each tweet and profile description was associated with the most probably topic. If a document belongs to different topics with low but similar probabilities such a hard assignment may not be nuanced enough. Thus, data on quantity (e.g., number of tweets falling within each topic) should be interpreted as indicative rather than definite. While we opted for LDA because of the human-interpretability of topics (cf.~\cite{Chang:2009}), others (e.g.,~\cite{Mehrotra:2013,Zhao:2011}) cautioned that LDA might not work well with Twitter and produce incoherent topics due to tweets being very short. This was, for example, evident with the topic labelled \textit{Various} (cf. Table~\ref{table:ldaTopicsTweets}) but by focusing on the topics with the highest coherence scores we could alleviate this issue in general. 

\section{Conclusions}

In this paper we explored a longitudinal dataset of tweets using unsupervised clustering and topic modelling to characterize the Twitter community formed around the multiplayer online shooter \emph{Destiny}. Our results indicate that there are distinct player groups with respect to their tweeting activity, but that Twitter-related features are not directly correlated with playtime. We identified several prevalent topics, including in-game content, streaming, infrastructure, news, and opinions. Further insights could be gained by looking into less common topics within the topic model. We also identified several subgroups of which the game's audience is composed of. However, apart from a few exceptions such as streamers and news outlets, these subgroups show a similar tweeting pattern with respect to the identified topics. While we feel that topic modelling can be an important tool for studying online communities, automatically extracted topics may not align with a specific framing of an issue an analyst may have in mind. In this sense, a combination of a human-driven visual analytics approach in combination with semi-supervised topic modelling techniques appears to be a promising direction for future work.

\section*{Acknowledgments}
Part of this work was conducted in the Digital Creativity Labs, jointly funded by EPSRC/AHRC/InnovateUK under grant no EP/M023265/1. We thank Christian Drescher and Rafet Sifa for compiling the dataset used in this research.

\balance
\bibliographystyle{IEEEtran}
\bibliography{references}

\end{document}